\documentclass[12pt]{article}
\usepackage{amsmath,amssymb,amsthm,amsxtra,overpic,bbm,bm,epsfig,subfigure}
\usepackage{hyperref}
\usepackage{mathrsfs}
\usepackage{enumitem}
\usepackage{graphicx}
\usepackage{color}
\usepackage{comment}
\usepackage{epstopdf}
\usepackage{float}
\usepackage{cite}
\usepackage{slashed,stmaryrd}

\def\thefootnote{\fnsymbol{footnote}}

\begin{document}

\vspace{0.2cm}

\begin{center}
{\Large\bf Towards the meV limit of the effective neutrino mass in neutrinoless double-beta decays}
\end{center}

\vspace{0.2cm}

\begin{center}
{\bf Jun Cao~$^{a,~b}$},
\quad
{\bf Guo-yuan Huang~$^{a,~b}$},
\quad
{\bf Yu-Feng Li~$^{a,~b}$},
\quad
{\bf Yifang Wang~$^{a,~b}$},
\quad
{\bf Liang-Jian Wen~$^{a,~b}$},
\quad
{\bf Zhi-zhong Xing~$^{a,~b}$},
\quad
{\bf Zhen-hua Zhao~$^{c}$},
\quad
{\bf Shun Zhou~$^{a,~b}$}~\footnote{E-mail: zhoush@ihep.ac.cn}
\\
\vspace{0.2cm}
{\small
$^a$Institute of High Energy Physics, Chinese Academy of Sciences, Beijing 100049, China\\
$^b$School of Physical Sciences, University of Chinese Academy of Sciences, Beijing 100049, China\\
$^c$Department of Physics, Liaoning Normal University, Dalian
116029, China \\}
\end{center}

\vspace{1.5cm}

\begin{abstract}
In this paper, we emphasize why it is important for future neutrinoless double-beta ($0\nu\beta\beta$) decay experiments to reach the sensitivity to the effective neutrino mass $|m^{}_{\beta\beta}| \approx 1~{\rm meV}$. Assuming such a sensitivity and the precisions on neutrino oscillation parameters after the JUNO experiment, we fully explore the constrained regions of the lightest neutrino mass $m^{}_1$ and two Majorana-type CP-violating phases $\{\rho, \sigma\}$. The implications for the neutrino mass spectrum, the effective neutrino mass $m^{}_\beta$ in beta decays and the sum of three neutrino masses $\Sigma \equiv m^{}_1 + m^{}_2 + m^{}_3$ relevant for cosmological observations are also discussed.
\end{abstract}


\def\thefootnote{\arabic{footnote}}
\setcounter{footnote}{0}
\newpage
\newpage
\section{Introduction}

Neutrino oscillation experiments have firmly established that neutrinos are massive particles and lepton flavors are significantly mixed~\cite{Wang:2015rma}. As expected in a class of seesaw models for neutrino mass generation~\cite{Xing:2011zza}, massive neutrinos are Majorana particles and they may provide a natural and elegant explanation for the observed matter-antimatter asymmetry in our Universe~\cite{Fukugita:1986hr}. If this is indeed the case, it will be a great challenge to determine two associated Majorana-type CP-violating phases, which are measurable only in the lepton-number-violating processes. Currently, the experimental search for neutrinoless double-beta ($0\nu\beta\beta$) decays ${^{A}_Z}N \to {^A_{Z+2}}N + 2e^-$ of some heavy nuclei ${^{A}_Z}N$, which possess an even atomic number $Z$ and an even mass number $A$, is the most promising way to demonstrate the Majorana nature of massive neutrinos and to prove the existence of lepton number violation in nature~\cite{Dolinski:2019nrj}. Assuming that three light Majorana neutrinos are responsible for the $0\nu\beta\beta$ decays of an even-even nuclear isotope, we can find the half-life~\cite{Bilenky:2014uka}
\begin{eqnarray}\label{eq:halflife}
T^{0\nu}_{1/2} = G^{-1}_{0\nu} \cdot \left|{\cal M}^{}_{0\nu}\right|^{-2} \cdot \left|m^{}_{\beta\beta}\right|^{-2} \cdot m^2_e \; ,
\end{eqnarray}
where $G^{}_{0\nu}$ denotes the relevant phase-space factor, ${\cal M}^{}_{0\nu}$ is the nuclear matrix element (NME), and $m^{}_e = 0.511~{\rm MeV}$ is the electron mass. In the standard parametrization of lepton flavor mixing matrix, the effective neutrino mass $|m^{}_{\beta \beta}|$ for $0\nu\beta\beta$ decays appearing in Eq.~(\ref{eq:halflife}) reads
\begin{eqnarray}\label{eq:mbb}
|m^{}_{\beta\beta}| \equiv \left|m^{}_1 \cos^2 \theta^{}_{13} \cos^2 \theta^{}_{12} e^{{\rm i}\rho} + m^{}_2 \cos^2 \theta^{}_{13} \sin^2 \theta^{}_{12} + m^{}_3 \sin^2 \theta^{}_{13} e^{{\rm i}\sigma}\right| \; ,
\end{eqnarray}
where $m^{}_i$ (for $i = 1, 2, 3$) stand for the absolute masses of three ordinary neutrinos. Out of three neutrino mixing angles only two $\{\theta^{}_{12}, \theta^{}_{13}\}$ are involved in the effective neutrino mass in Eq.~(\ref{eq:mbb}), where $\{\rho, \sigma\}$ are two Majorana CP phases. The other neutrino mixing angle $\theta^{}_{23}$ and the Dirac-type CP-violating phase $\delta$ are irrelevant for $0\nu\beta\beta$ decays.

The main purpose of the present study is to explore the physics potential of pinning down the fundamental parameters in future $0\nu\beta\beta$ decay experiments, in particular the absolute neutrino masses and two Majorana phases that are not accessible at all in neutrino oscillation experiments. Our motivation is three-fold:
\begin{itemize}
\item Neutrino oscillation experiments have measured with reasonably good precisions the relevant two neutrino mixing angles $\{\theta^{}_{12}, \theta^{}_{13}\}$, and two independent neutrino mass-squared differences $\Delta m^2_{21} \equiv m^2_2 - m^2_1$ and $|\Delta m^2_{31}| \equiv |m^2_3 - m^2_1|$~\cite{Wang:2015rma}. In the near future, the JUNO experiment~\cite{Li:2013zyd, An:2015jdp} will be able to offer an unambiguous answer to whether neutrino mass ordering is normal $m^{}_1 < m^{}_2 < m^{}_3$ (NO) or inverted $m^{}_3 < m^{}_1 < m^{}_2$ (IO), and to improve the precisions of three parameters $\{\sin^2 \theta^{}_{12}, \Delta m^2_{21}, \Delta m^2_{31}\}$ to the level below one percent. In addition, the ultimate precision on $\sin^2\theta^{}_{13}$ from the Daya Bay experiment will be $3\%$~\cite{Vogel:2015wua, Cao:2017drk}. According to neutrino oscillation data, at least two neutrino masses should be above the ${\rm meV}$ level, e.g., $m^{}_3 > m^{}_2 = \sqrt{m^2_1 + \Delta m^2_{21}} \geq \sqrt{\Delta m^2_{21}} \approx 8.6~{\rm meV}$ (for NO). Given oscillation parameters, the observation of $0\nu\beta\beta$ decays will be extremely important in the determination of the lightest neutrino mass $m^{}_{\rm 1}$ (for NO) or $m^{}_3$ (for IO) and two Majorana CP phases $\{\rho, \sigma\}$.

\item The upper bound on the absolute scale of neutrino masses extracted from the tritium beta decays is $m^{}_\beta \leq 2.3~{\rm eV}$ (Mainz~\cite{Kraus:2004zw}) and $m^{}_\beta \leq 2.2~{\rm eV}$ (Troitsk~\cite{Aseev:2011dq}) at the $95\%$ confidence level, where the effective neutrino mass $m^{}_\beta$ for beta decays is defined as $m^{}_\beta \equiv \left( m^2_1 |U^{}_{e1}|^2 + m^2_2 |U^{}_{e2}|^2 + m^2_3 |U^{}_{e3}|^2\right)^{1/2}$ with the moduli of the matrix elements of lepton flavor mixing matrix being $|U^{}_{e1}| = \cos \theta^{}_{13} \cos \theta^{}_{12}$, $|U^{}_{e2}| = \cos \theta^{}_{13} \sin \theta^{}_{12}$ and $|U^{}_{e3}| = \sin \theta^{}_{13}$. The next-generation tritium beta-decay experiments KATRIN~\cite{Osipowicz:2001sq, Wolf:2008hf} and Project 8~\cite{Esfahani:2017dmu} will hopefully be capable of bringing the upper limit down to $m^{}_\beta \leq 200~{\rm meV}$ and $m^{}_\beta \leq 40~{\rm meV}$, respectively. On the other hand, the cosmological observations of cosmic microwave background by the Planck satellite gives the most restrictive bound on the sum of three neutrino masses $\Sigma \equiv m^{}_1 + m^{}_2 + m^{}_3 < 120~{\rm meV}$~\cite{Aghanim:2018eyx}. However, there is still a long way to go until the neutrino mass region of a few meV is accessed.

\item Future large and ultra-low background liquid scintillator (LS) detectors, such as JUNO, have great potential of searching for $0\nu\beta\beta$ decays, by dissolving the $0\nu\beta\beta$-decaying isotope $^{130}{\rm Te}$ or $^{136}{\rm Xe}$ into LS. This concept has been discussed in Ref.~\cite{Zhao:2016brs}, where the Xe-loaded LS is taken as a target. It has been demonstrated that a sensitivity (at the $90\%$ confidence level) to $T^{1/2}_{0\nu}$ of $1.8 \times 10^{28}~{\rm yr}$ is achievable with 50 tons of fiducial $^{136}{\rm Xe}$ and 5 years of exposure, while the corresponding sensitivity to the effective neutrino mass $|m^{}_{\beta\beta}|$ could reach $(5\cdots 12)~{\rm meV}$ depending on the NME value. It has also been pointed out that $^{130}{\rm Te}$ may be an advantageous candidate due to its high natural abundance. If the nuclear isotope $^{130}{\rm Te}$ with a maximum fraction of $4\%$ is loaded, a total target mass of $400$ tons could be obtained, leading to an improvement on the sensitivity by a factor of $(400/50)^{1/4} \approx 1.68$, namely, $|m^{}_{\beta\beta}| \approx (2.3\cdots 6.0)~{\rm meV}$ for the same background index.\footnote{Note that the sensitivity to $|m^{}_{\beta\beta}|$ depends on the relevant NME value, given the experimental setups with the same target mass, exposure and background index. Here the NME values of the $0\nu\beta\beta$ decays of ${^{130}}{\rm Te}$ have been taken from the same references for those of ${^{136}}{\rm Xe}$ as in Ref.~\cite{Zhao:2016brs} such that both NME values are calculated in the same theoretical nuclear model. In all the relevant theoretical models, the NME values for $^{130}{\rm Te}$ are larger than those for $^{136}{\rm Xe}$, which is another advantage of the $^{130}{\rm Te}$ option. If the target mass $M$ is increased, we estimate the improved sensitivity to $|m^{}_{\beta\beta}|$ by following the scaling law $|m^{}_{\beta\beta}| \propto M^{1/4}$~\cite{Dolinski:2019nrj}. } Moreover, if the nominal value of the background index in Ref.~\cite{Zhao:2016brs} is further significantly reduced, e.g., by two orders of magnitude, the ultimate sensitivity will hopefully be close to $|m^{}_{\beta\beta}| \approx 1~{\rm meV}$.
\end{itemize}

In the literature, it has been noticed~\cite{XZZ, XZ, XZ2, Ge:2016tfx, Penedo:2018kpc} that $|m^{}_{\beta\beta}| \approx 1~{\rm meV}$ could be set as a target value and useful information on the absolute neutrino masses and the Majorana CP phases can be obtained. The present study differs from previous works in two aspects. First, we concentrate on the effective neutrino mass $|m^{}_{\beta\beta}|$ in the NO case and update its value with both the latest global-fit results of all the relevant neutrino oscillation parameters and the future measurements from neutrino oscillation experiments. With these input, it becomes clearer how much and definite the effective mass $|m^{}_{\beta\beta}| \approx 1~{\rm meV}$ can tell us the information about the lightest neutrino mass $m^{}_1$ and the Majorana phases $\{\rho, \sigma\}$. Second, we further explore the implications for the neutrino mass spectrum, the effective mass $m^{}_\beta$ for beta decays and the sum of three neutrino masses $\Sigma$, and stress that the determination of the lightest neutrino mass from the $0\nu\beta\beta$ decay experiment with a sensitivity of $|m^{}_{\beta\beta}| \approx 1~{\rm meV}$ sets up a challenging goal for future beta-decay experiments and cosmological observations.

The remaining part of this paper is organized as follows. In Sec.~\ref{sec:effmass}, the three-dimensional description of the effective neutrino mass $|m^{}_{\beta\beta}|$ as a function of the lightest neutrino mass $m^{}_1$ and the Majorana CP phase $\rho$ is given, where the future precision on neutrino oscillation parameters is implemented and the latest global-fit results from Ref.~\cite{Esteban:2018azc} are also considered for comparison. Furthermore, the implications for the neutrino mass spectrum, the effective neutrino mass $m^{}_\beta$ in beta decays and the sum of three neutrino masses from cosmological observations are discussed in Sec.~\ref{sec:beta}. Finally, we give some further remarks and summarize our main conclusions in Sec.~\ref{sec:conc}.

\section{Neutrino Masses and Majorana CP Phases}\label{sec:effmass}

\subsection{Two-dimensional description}

\begin{figure}[t!]
	\begin{center}
		\subfigure{
			\includegraphics[width=0.47\textwidth]{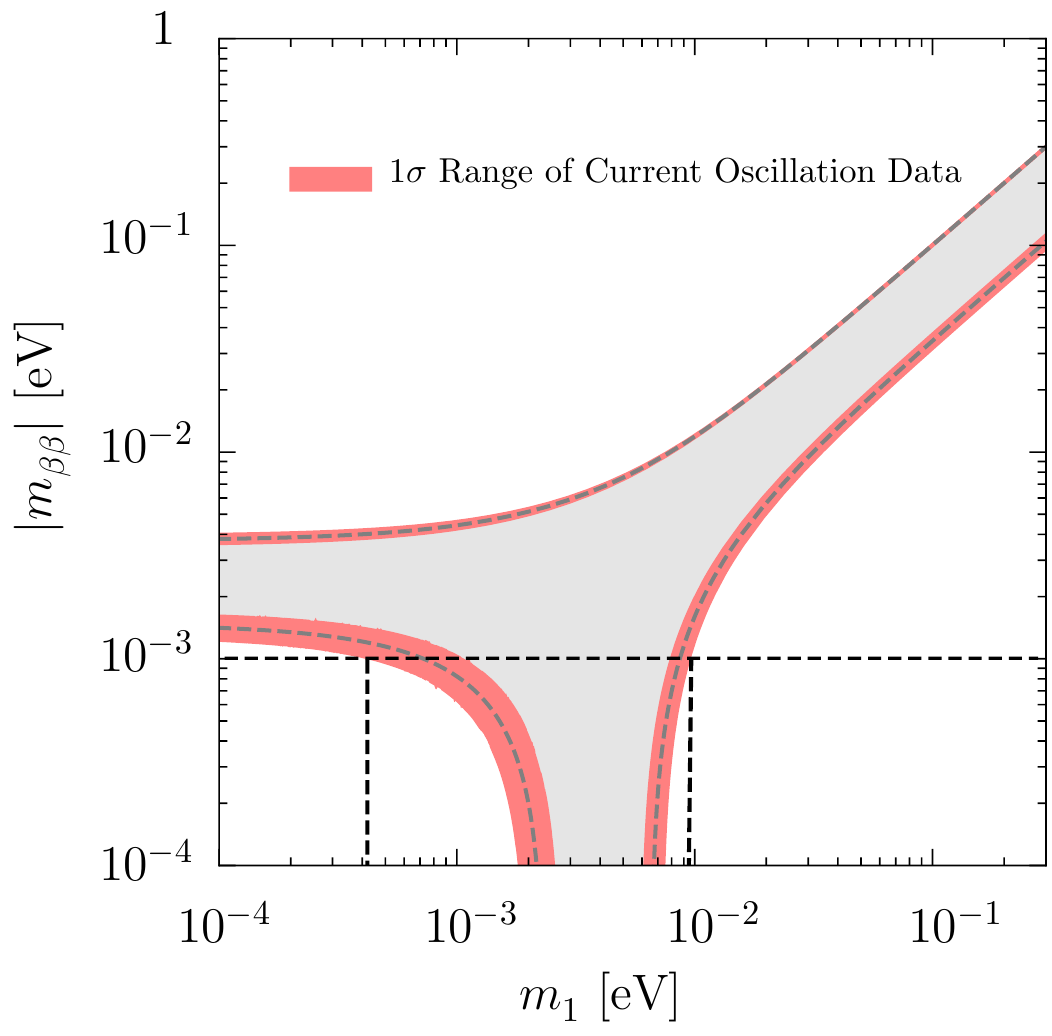} }
		\subfigure{
			\includegraphics[width=0.47\textwidth]{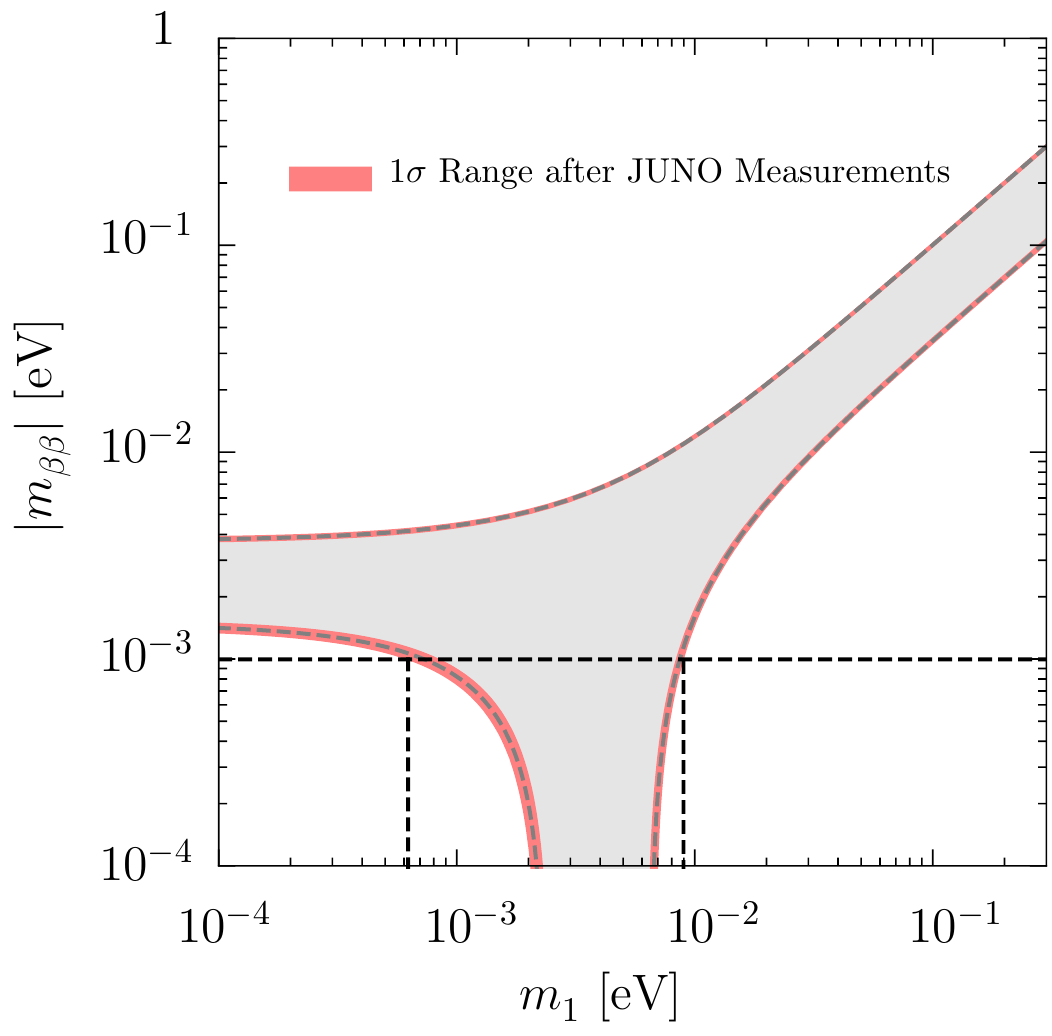} }
	\end{center}
	\vspace{-0.5cm}
	\caption{The effective neutrino mass $|m^{}_{\beta\beta}|$ is shown as a function of the lightest neutrino mass $m^{}_1$ in the NO case, where the gray region is allowed and the dashed curves refer to its boundaries. In the left panel, the boundaries are obtained by using the best-fit values of $\{\theta^{}_{12}, \theta^{}_{13}\}$ and $\{\Delta m^2_{21}, \Delta m^2_{31}\}$ from the latest global-fit analysis of neutrino oscillation parameters in Ref.~\cite{Esteban:2018azc} and the red region is caused by the $1\sigma$ uncertainties of these parameters. In the right panel, the future precisions on those oscillation parameters after the JUNO measurements~\cite{An:2015jdp} are implemented. In each panel, the horizontal dashed line corresponds to $|m^{}_{\beta\beta}| = 1~{\rm meV}$ and the allowed range of $m^{}_1$ is indicated by two vertical dashed lines.}
	\label{fig:Vissani}
\end{figure}

The conventional way of graphically showing the possible range of $|m^{}_{\beta\beta}|$ is to plot it as a function of the lightest neutrino mass ($m^{}_1$ for the NO case or $m^{}_3$ for the IO case) by varying $\rho$ and $\sigma$ in the whole range of $[0, 360^\circ)$, as first suggested in Ref.~\cite{Vissani}. In Fig.~\ref{fig:Vissani}, the allowed range of $|m^{}_{\beta\beta}|$ in the NO case is shown as the gray region. The boundaries of the allowed range are denoted by the dashed curves, which are obtained by using the best-fit values of $\{\theta^{}_{12}, \theta^{}_{13}\}$ and $\{\Delta m^2_{21}, \Delta m^2_{31}\}$ from the latest global-fit analysis of neutrino oscillation data in Ref.~\cite{Esteban:2018azc}. In the left panel, the red bands along the dashed curves are caused by the $1\sigma$ uncertainties of the oscillation parameters from the global-fit analysis, while those in the right panel are due to the $1\sigma$ uncertainties after the JUNO measurements~\cite{An:2015jdp}. In each panel, the horizontal dashed line corresponds to $|m^{}_{\beta\beta}| = 1~{\rm meV}$ and the allowed range of $m^{}_1$ is indicated by two vertical dashed lines. More explicitly, we quote the best-fit values and current uncertainties of the relevant oscillation parameters from Ref.~\cite{Esteban:2018azc} in the NO case as below
\begin{eqnarray}\label{eq:data}
\begin{array}{lcl}
  \sin^2 \theta^{}_{12} = 0.310^{+0.013}_{-0.012} \; , & \quad & \Delta m^2_{21} = 7.39^{+0.21}_{-0.20} \times 10^{-5}~{\rm eV}^2 \; ; \\
  \sin^2 \theta^{}_{13} = 0.02241^{+0.00066}_{-0.00065} \; , & \quad & \Delta m^2_{31} = 2.523^{+0.032}_{-0.030} \times 10^{-3}~{\rm eV}^2 \; .
\end{array}
\end{eqnarray}
For the future measurements of these parameters, we assume that the best-fit values are the same, but the precision on $\sin^2 \theta^{}_{12}$, $\Delta m^2_{21}$ and $\Delta m^2_{31}$ will be improved after the JUNO experiment to $0.54\%$, $0.24\%$ and $0.27\%$, respectively. It is worth mentioning that the ultimate precision of $3\%$ on $\sin^2\theta^{}_{13}$ from the Daya Bay experiment will be adopted, which is comparable to that in Eq.~(\ref{eq:data}).

In order to clarify the dependence of $|m^{}_{\beta\beta}|$ on the oscillation parameters, we consider its upper (``U") and lower (``L") boundaries that are derived by varying the Majorana CP phase $\sigma$, namely,
\begin{eqnarray}\label{eq:UL}
|m^{}_{\beta \beta}|^{}_{\rm U, L} \equiv ||\overline{m}^{}_{12}| \pm m^{}_3 \sin^2 \theta^{}_{13}| \; ,
\end{eqnarray}
where the sign ``$+$" (or ``$-$") corresponds to ``U" (or ``L"), and $\overline{m}^{}_{12} \equiv m^{}_1 \cos^2 \theta^{}_{13} \cos^2 \theta^{}_{12} e^{{\rm i}\rho} + m^{}_2 \cos^2 \theta^{}_{13} \sin^2 \theta^{}_{12}$ is the sum of the first two terms in $m^{}_{\beta\beta}$ defined in Eq.~(\ref{eq:mbb}). Notice that the phase $\sigma$ has been properly chosen to match exactly (or differ by $\pm \pi$ from) the phase of $\overline{m}^{}_{12}$ to draw the upper (lower) boundary. Some comments on the upper and lower boundaries of $|m^{}_{\beta\beta}|$ in Fig.~\ref{fig:Vissani} are in order.

In the NO case, given two neutrino mass-squared differences $\Delta m^2_{21} \simeq 7.39\times 10^{-5}~{\rm eV}^2$ and $\Delta m^2_{31} \simeq 2.523\times 10^{-3}~{\rm eV}^2$, we have $m^{}_3 = \sqrt{m^2_1 + \Delta m^2_{31}}$ and $m^{}_2 = \sqrt{m^2_1 + \Delta m^2_{21}}$. The upper boundary is determined by
\begin{eqnarray}\label{eq:NOUB}
|m^{}_{\beta\beta}|^{}_{\rm U} = \cos^2 \theta^{}_{13} \left(m^{}_1 \cos^2\theta^{}_{12} + \sin^2 \theta^{}_{12} \sqrt{\Delta m^2_{21} + m^2_1}\right) + \sin^2 \theta^{}_{13} \sqrt{\Delta m^2_{31} + m^2_1}  \; ,
\end{eqnarray}
implying $|m^{}_{\beta\beta}|^{}_{\rm U} \to \sin^2 \theta^{}_{12} \cos^2\theta^{}_{13} \sqrt{\Delta m^2_{21}} + \sin^2\theta^{}_{13} \sqrt{\Delta m^2_{31}}$ in the limit of $m^{}_1 \to 0$. In this limit, the first term $\sin^2 \theta^{}_{12} \cos^2 \theta^{}_{13} \sqrt{\Delta m^2_{21}} \simeq 2.6~{\rm meV}$ is about twice larger than the second term $\sin^2\theta^{}_{13} \sqrt{\Delta m^2_{31}}\simeq 1.1~{\rm meV}$, where $\sin^2 \theta^{}_{12} \simeq 0.310$ and $\sin^2 \theta^{}_{13} \simeq 0.02241$ have been used. Hence the uncertainties from $\sin^2 \theta^{}_{12}$ and $\Delta m^2_{21}$ dominate over those from $\sin^2 \theta^{}_{13}$ and $\Delta m^2_{31}$. This can be clearly seen by comparing the uncertainty of the upper boundary in the left panel of Fig.~\ref{fig:Vissani} with that in the right panel, as the improvements on $\sin^2 \theta^{}_{12}$ and $\Delta m^2_{21}$ after the JUNO measurements are remarkable. When the lightest neutrino mass $m^{}_1$ increases, in particular for $m^{}_1 \gtrsim \sqrt{\Delta m^2_{21}} \simeq 8.6~{\rm meV}$, we obtain $m^{}_1 \simeq m^{}_2$ and the first term on the right-hand side of Eq.~(\ref{eq:NOUB}) is approximately given by $m^{}_1 \cos^2 \theta^{}_{13}$, indicating that the dependence on the $\sin^2 \theta^{}_{12}$ and $\Delta m^2_{21}$ becomes very weak. On the other hand, the lower boundary in the limit of $m^{}_1 \to 0$ reads
\begin{eqnarray}\label{eq:NOLB}
|m^{}_{\beta\beta}|^{}_{\rm L} = \cos^2 \theta^{}_{13} \sin^2 \theta^{}_{12} \sqrt{\Delta m^2_{21}} - \sin^2 \theta^{}_{13} \sqrt{\Delta m^2_{31}} \; ,
\end{eqnarray}
which is estimated to be $|m^{}_{\beta\beta}|^{}_{\rm L} \simeq 1.5~{\rm meV}$ for the best-fit values of neutrino mixing angles and mass-squared differences in Eq.~(\ref{eq:data}). Since the first term on the right-hand side of Eq.~(\ref{eq:NOLB}) is about twice larger than the second term, the dominant uncertainty on the lower boundary comes from $\sin^2 \theta^{}_{12}$ and $\Delta m^2_{21}$. For $m^{}_1 \gtrsim \sqrt{\Delta m^2_{21}} \simeq 8.6~{\rm meV}$ and thus $m^{}_1\simeq m^{}_2$, we obtain $|m^{}_{\beta\beta}|^{}_{\rm L} \simeq \cos^2\theta^{}_{13} \cos2\theta^{}_{12} \sqrt{\Delta m^2_{21} + m^2_1} - \sin^2\theta^{}_{13} \sqrt{\Delta m^2_{31} + m^2_1}$, so the uncertainty from $\sin^2 \theta^{}_{12}$ continues to be dominant as clearly shown in Fig.~\ref{fig:Vissani}. In the region of $2~{\rm meV} \lesssim m^{}_1 \lesssim 7~{\rm meV}$, the destructive cancellation appears in $|m^{}_{\beta\beta}|$ due to the Majorana CP phases, leading to a ``well"-like structure, which will be examined more carefully in the next subsection.

We have observed that the boundaries of the allowed range of $|m^{}_{\beta\beta}|$ in the NO case depend crucially on the precision of $\sin^2 \theta^{}_{12}$ and $\Delta m^2_{21}$. After the JUNO measurements, as illustrated in the right panel of Fig.~\ref{fig:Vissani}, the uncertainties from neutrino mixing angles and mass-squared differences could be safely ignored.

\subsection{Three-dimensional description}

As we have mentioned, in the so-called Vissani graph~\cite{Vissani} in Fig.~\ref{fig:Vissani}, there is a ``well"-like structure of $|m^{}_{\beta\beta}|$ in the region of $2~{\rm meV} \lesssim m^{}_1 \lesssim 7~{\rm meV}$ for the NO case. Inside the well, $|m^{}_{\beta\beta}|$ takes tiny values, where a significant cancellation among three components of $m^{}_{\beta\beta}$ occurs. The bottom of the well signifies the extreme case of $|m^{}_{\beta\beta}| \to 0$, where a complete cancellation takes place. In Ref.~\cite{XZZ}, the three-dimensional graph, where $|m^{}_{\beta\beta}|$ is plotted against both the lightest neutrino mass $m^{}_1$ and the Majorana CP phase $\rho$, has been suggested and shown to be extremely useful in revealing the fine structure inside the well~\cite{XZ, XZ2}.
\begin{figure}[!t]
\centering
\includegraphics[width=0.88\textwidth]{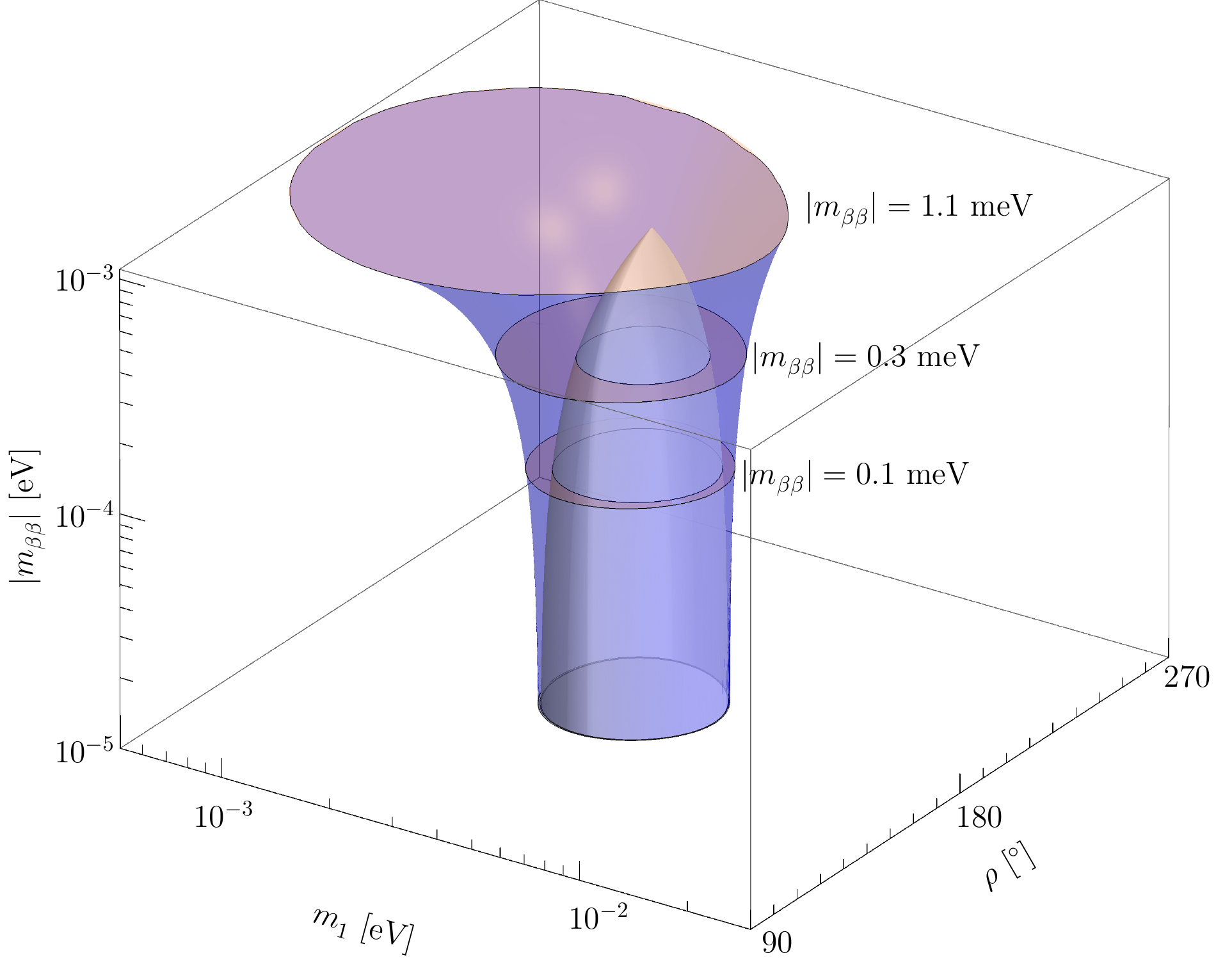}
\vspace{-0.2cm}
\caption{The zoomed-in three-dimensional graph of $|m^{}_{\beta\beta}|$ as a function $m^{}_1$ and $\rho$ for the NO case, where the best-fit values $\Delta m^2_{21} = 7.39 \times 10^{-5}~{\rm eV}^2$, $\Delta m^2_{31} = 2.523 \times 10^{-3}~{\rm eV}^2$, $\sin^2\theta^{}_{12} = 0.310$ and $\sin^2\theta^{}_{13} = 0.02241$ have been taken~\cite{Esteban:2018azc}. The blue filled volume stands for the allowed range of $|m^{}_{\beta\beta}|$, while three contour surfaces corresponding to $|m^{}_{\beta\beta}| = 1.1~{\rm meV}$, $0.3~{\rm meV}$ and $0.1~{\rm meV}$ are shown and the ``bullet"-like region is hollowed out.}
\label{fig:3dUL}
\end{figure}

In Fig.~\ref{fig:3dUL}, we reproduce the three-dimensional graph of $|m^{}_{\beta\beta}|$ in the NO case from Refs.~\cite{XZ, XZ2} and zoom in the region of $10^{-2}~{\rm meV} \leq |m^{}_{\beta\beta}| \leq 1.1~{\rm meV}$, where the best-fit values of neutrino mixing angles and mass-squared differences from Ref.~\cite{Esteban:2018azc} have been used. As we have explained, the upper (or lower) boundary of the allowed range of $|m^{}_{\beta\beta}|$ can be obtained by properly choosing the Majorana CP phase $\sigma$ such that it matches (or differs by $\pm \pi$ from) the phase of $\overline{m}^{}_{12}$. More explicitly, the values of $\sigma$ are determined by
\begin{eqnarray}\label{eq:sigma}
\sin\sigma & = & \pm \frac{m^{}_1 \sin\rho}{\sqrt{m^2_1+2 m^{}_1 m^{}_2 \tan^2 \theta^{}_{12} \cos \rho + m^2_2 \tan^4 \theta^{}_{12} }} \;, \nonumber \\
\cos\sigma & = & \pm \frac{m^{}_1 \cos \rho + m^{}_2 \tan^2\theta^{}_{12} }{\sqrt{m^2_1+2 m^{}_1 m^{}_2 \tan^2\theta^{}_{12} \cos \rho + m^2_2 \tan^4\theta^{}_{12} }} \;.
\end{eqnarray}
With the help of Eq.~(\ref{eq:UL}), one finds that the bottom of the well (i.e., $|m^{}_{\beta\beta}|^{}_{\rm L} = 0$) would be reached for $
\left| \overline{m}^{}_{12} \right| = m^{}_3 \sin^2\theta^{}_{13}$. Then it is straightforward to verify that $|m^{}_{\beta\beta}| = 0$ holds only in the narrow region of $2 ~{\rm meV} \lesssim m^{}_1 \lesssim 7 ~{\rm meV}$ and $155^\circ \lesssim \rho \lesssim 205^\circ$. The other Majorana CP phase $\sigma$ is fixed via Eq.~(\ref{eq:sigma}) by choosing the minus sign.

As shown in Fig.~\ref{fig:3dUL}, three contour surfaces corresponding to $|m^{}_{\beta\beta}| = 1.1~{\rm meV}$, $0.3~{\rm meV}$ and $0.1~{\rm meV}$ appear above the ``bullet"-like structure. The surface of this bullet is described by
\begin{eqnarray}\label{eq:bullet}
|m^{}_{\beta\beta}|^{}_{\rm L} = m^{}_3 \sin^2\theta^{}_{13}
- |\overline{m}^{}_{12}|  \;,
\end{eqnarray}
whose maximum $|m^{}_{\beta\beta}|^{}_{*} = m^{}_3 \sin^2 \theta^{}_{13}$ appears at $\overline{m}^{}_{12} = 0$ that in turn requires $\rho = 180^\circ$ and $m^{}_1/m^{}_2 = \tan^2\theta^{}_{12}$~\cite{XZ}. Numerically, for the best-fit values of oscillation parameters in Eq.~(\ref{eq:data}), the tip of the bullet is located at $\left(m^{}_1, \rho, |m^{}_{\beta\beta}|^{}_{*}\right) \simeq \left(4 ~ {\rm meV}, 180^\circ, 1.1~{\rm meV}\right)$. Note that the region covered by the bullet surface is actually hollowed out, so the allowed parameter space for small values of $|m^{}_{\beta\beta}|$ decreases rapidly, which becomes clear by comparing among three contour surfaces in Fig.~\ref{fig:3dUL}. For the following two reasons, $|m^{}_{\beta\beta}|^{}_{*} \simeq 1.1~{\rm meV}$ can be taken as a threshold value in some sense. On the one hand, as one has already seen, the parameter space for $|m^{}_{\beta\beta}| \lesssim |m^{}_{\beta\beta}|^{}_{*}$ to hold is very small as compared with the whole parameter space. On the other hand, although it is obviously challenging for the future $0\nu\beta\beta$ decay experiments to reach the sensitivity of $|m^{}_{\beta\beta}| \approx 1~{\rm meV}$, this goal is hopefully achievable by further optimizing and improving the experimental setup considered in Ref.~\cite{Zhao:2016brs}. If such a sensitivity is ultimately reached, one can probe the absolute neutrino masses to an unprecedented precision at the meV level and draw a restrictive constraint on the Majorana CP phases.

\begin{figure}[t!]
\begin{center}
\includegraphics[width=0.75\textwidth]{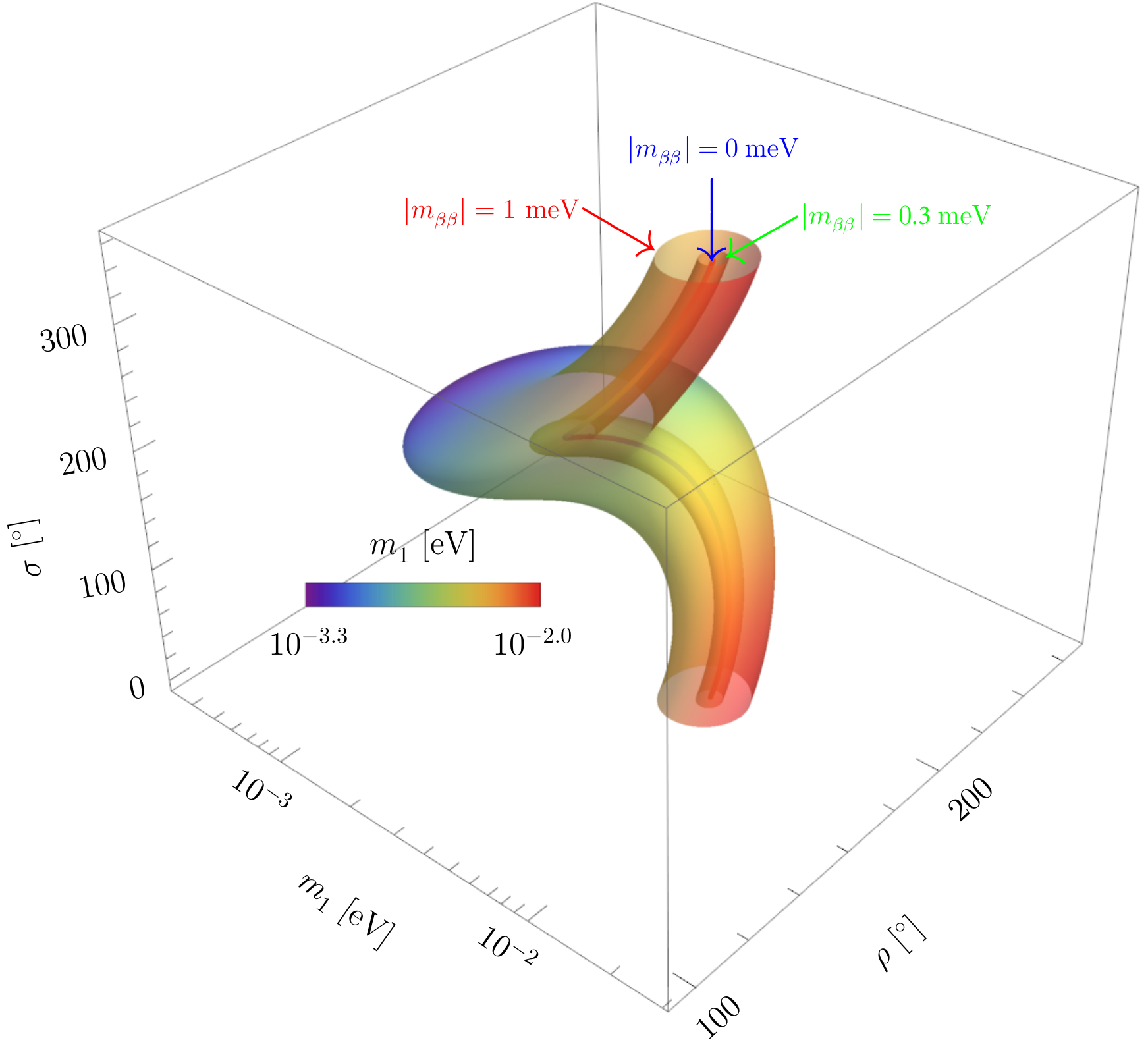}\\
\includegraphics[width=\textwidth]{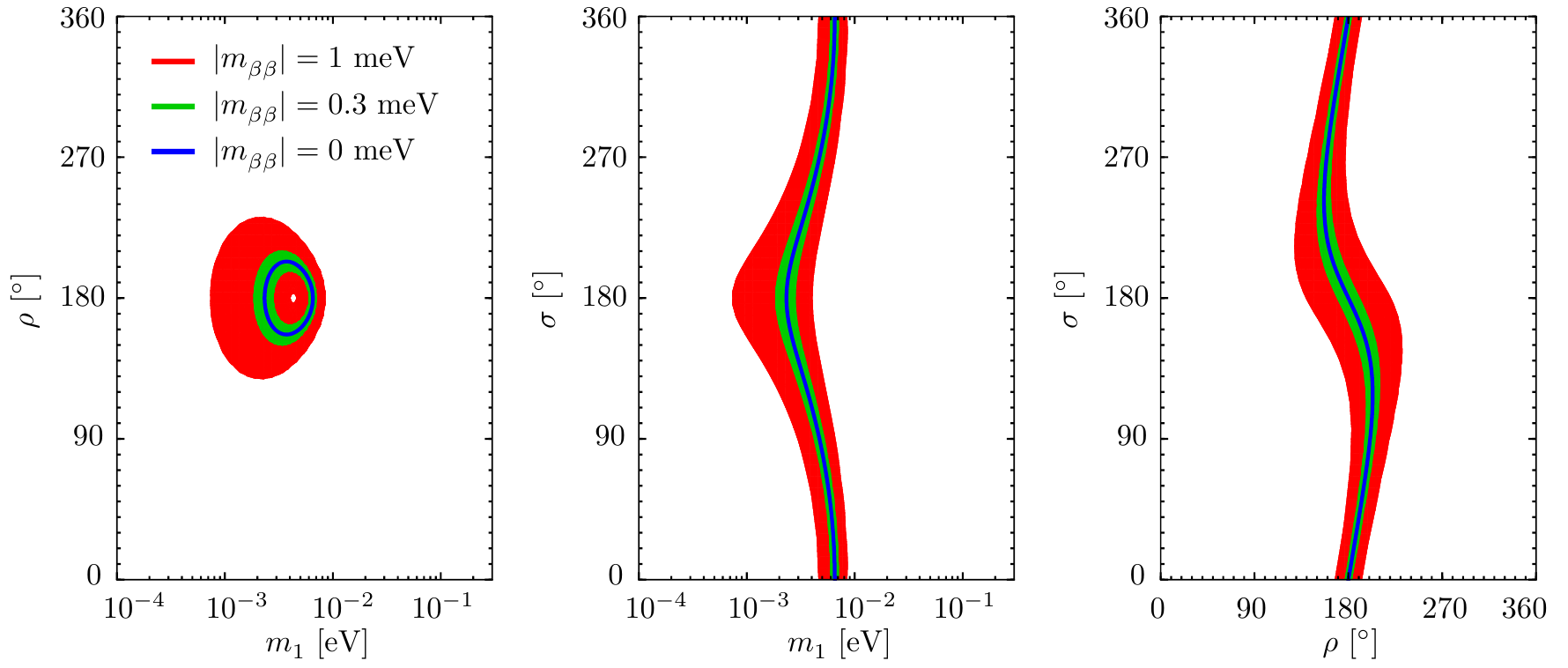}
\end{center}
\vspace{-0.5cm}
\caption{The allowed parameter space of $m^{}_1$, $\rho$ and $\sigma$ for $|m^{}_{\beta\beta}| \leq 1~{\rm meV}$, $|m^{}_{\beta\beta}| \leq 0.3~{\rm meV}$ and $|m^{}_{\beta\beta}| = 0$, where the same best-fit values of $\Delta m^2_{21}$, $\Delta m^2_{31}$, $\sin^2\theta^{}_{12}$ and $\sin^2\theta^{}_{13}$ as in Fig.~\ref{fig:3dUL} have been used. In the lower panel, the projections of the allowed parameter space into the $(m^{}_1, \rho)$-, $(m^{}_1, \sigma)$- and $(\rho, \sigma)$-plane have also been shown.}
\label{fig:3dmrs}
\end{figure}

Now let us assume that the sensitivity of $|m^{}_{\beta\beta}| = 1~{\rm meV}$ or even below can be reached in the future $0\nu\beta\beta$ decay experiments. In this case, the allowed parameter space of $(m^{}_1, \rho, \sigma)$ is given in Fig.~\ref{fig:3dmrs}. For comparison, we have shown the results for $|m^{}_{\beta\beta}| = 1~{\rm meV}$, $0.3~{\rm meV}$ and $0~{\rm meV}$ in the upper panel. In the lower panel, the projections of the parameter space into the $(m^{}_1, \rho)$-, $(m^{}_1, \sigma)$- and $(\rho, \sigma)$-plane have also been presented. Some comments on the numerical results are helpful.
\begin{itemize}
\item In our calculations, the best-fit values of neutrino mixing angles and mass-squared differences in Eq.~(\ref{eq:data}) have been adopted. After taking account of the JUNO measurements, the uncertainties of those parameters will be negligible. In fact, we have also numerically checked that this is indeed true.

\item If the sensitivity of $|m^{}_{\beta\beta}| = 1~{\rm meV}$ is eventually achieved, the allowed ranges of three fundamental parameters $m^{}_1$, $\rho$ and $\sigma$ can be read off from the red shaded areas in the two-dimensional plots in the lower panel of Fig.~\ref{fig:3dmrs}. More explicitly, we have $m^{}_1 \in [0.7, 8]~{\rm meV}$, $\rho \in [130^\circ, 230^\circ]$ and $\sigma \in [0, 360^\circ)$. Note that there is a tiny blank region in the center of the red area in the $(m^{}_1, \rho)$-plane. This is due to the fact that the tip of the bullet is located at $|m^{}_{\beta\beta}| = |m^{}_{\beta\beta}|^{}_* \simeq 1.1~{\rm meV}$, which is slightly above the contour surface of $|m^{}_{\beta\beta}| = 1~{\rm meV}$.

\item When the sensitivity is further improved to $|m^{}_{\beta\beta}| = 0.3~{\rm meV}$, only the green bands in the two-dimensional plots are allowed. As a consequence, the parameter space turns out to be more strictly constrained, namely, $m^{}_1 \in [1, 7]~{\rm meV}$ and $\rho \in [150^\circ, 210^\circ]$. However, the whole range of $\sigma$ is still allowed. In the extreme case of $|m^{}_{\beta\beta}| = 0~{\rm meV}$, the parameter space is represented by the blue curves. Given the lightest neutrino mass $m^{}_1$, one can completely pin down the values of the Majorana CP phases $\rho$ and $\sigma$. Even in this case, depending on the exact value of $m^{}_1$, any value of $\sigma$ within $[0, 360^\circ)$ can be taken.
\end{itemize}

\begin{figure}[t!]
\begin{center}
\includegraphics[width=0.6\textwidth]{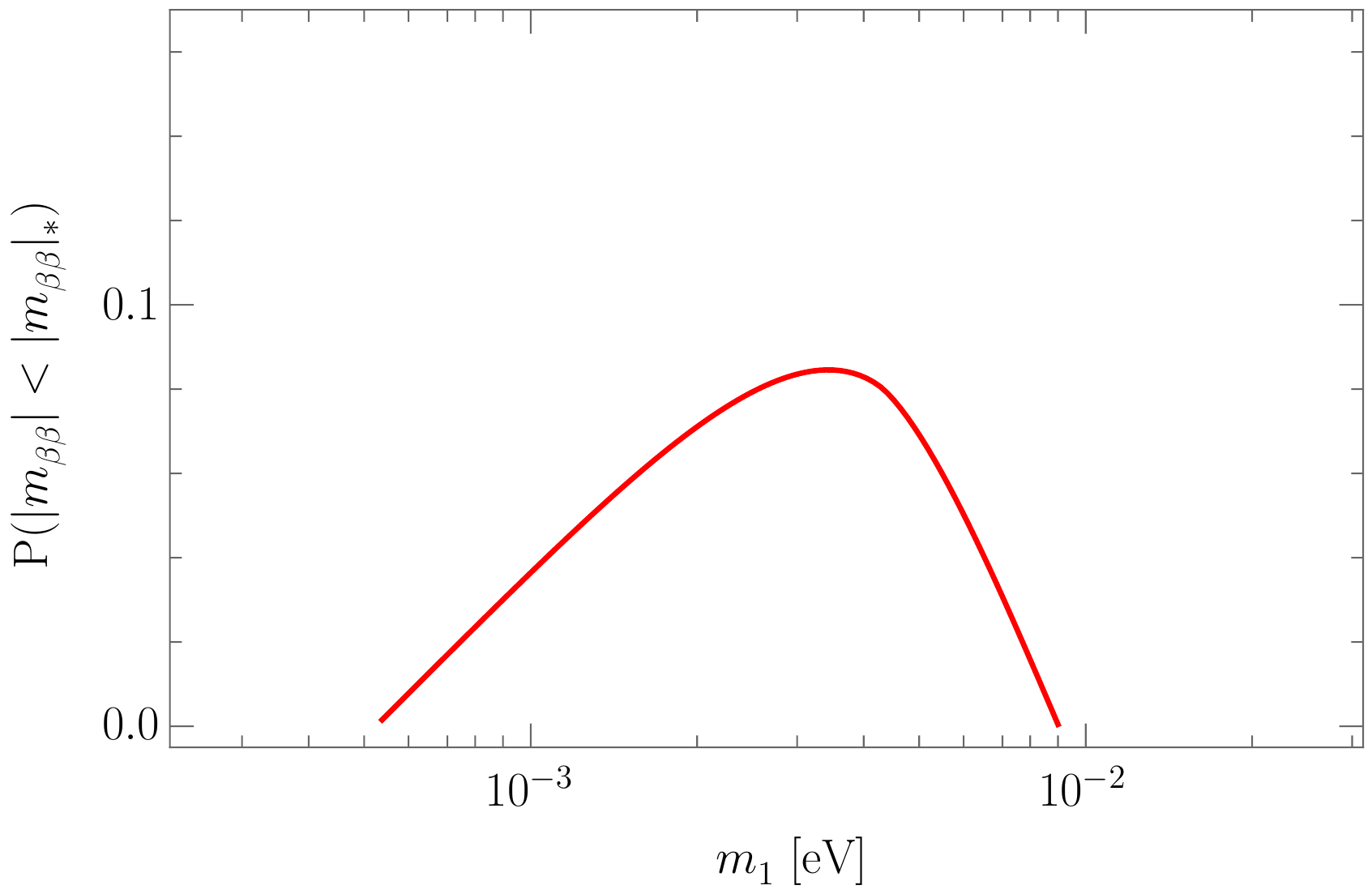}
\end{center}
\vspace{-0.5cm}
\caption{The ratio of the region of $(\rho, \sigma)$ for $|m^{}_{\beta\beta}| < |m^{}_{\beta\beta}|^{}_{*}$ to hold to the whole region $360^\circ \times 360^\circ$ for any given value of $m^{}_1$, where the best-fit values of $\Delta m^2_{21}$, $\Delta m^2_{31}$, $\sin^2\theta^{}_{12}$ and $\sin^2\theta^{}_{13}$ have been input as in Fig.~\ref{fig:3dUL}.}
\label{fig:sigmafrac}
\end{figure}

To quantitatively explain the constraining power of the $0\nu\beta\beta$ decay experiments, we introduce the probability $P(|m^{}_{\beta\beta}| < |m^{}_{\beta\beta}|^{}_*)$ as a function of the lightest neutrino mass $m^{}_1$. For a given value of $m^{}_1$, this probability is calculated as the ratio of the required ranges of $\rho$ and $\sigma$ (for $|m^{}_{\beta\beta}| < |m^{}_{\beta\beta}|^{}_* \simeq 1.1~{\rm meV}$ to be satisfied) to the whole range of $360^\circ \times 360^\circ$. In Fig.~\ref{fig:sigmafrac}, $P(|m^{}_{\beta\beta}| < |m^{}_{\beta\beta}|^{}_*)$ has been plotted as the red solid curve. One can observe that this fraction is always below $10\%$ for any viable value of $m^{}_1$. Based on the above calculations, we conclude that the fulfillment of $|m^{}_{\beta\beta}| \lesssim |m^{}_{\beta\beta}|^{}_* \simeq 1.1~{\rm meV}$ requires significant cancellation among three terms in $m^{}_{\beta\beta}$ and the possibility for such a case to come true is really small~\cite{XZ, XZ2}. In other words, future $0\nu\beta\beta$ experiments with a sensitivity of $|m^{}_{\beta\beta}| \approx 1~{\rm meV}$ will be able to determine the lightest neutrino mass with a high precision and even to probe the Majorana CP phases.

\section{Implications for Beta Decays and Cosmology}\label{sec:beta}

\begin{figure}[t!]
\begin{center}
\includegraphics[width=0.6\textwidth]{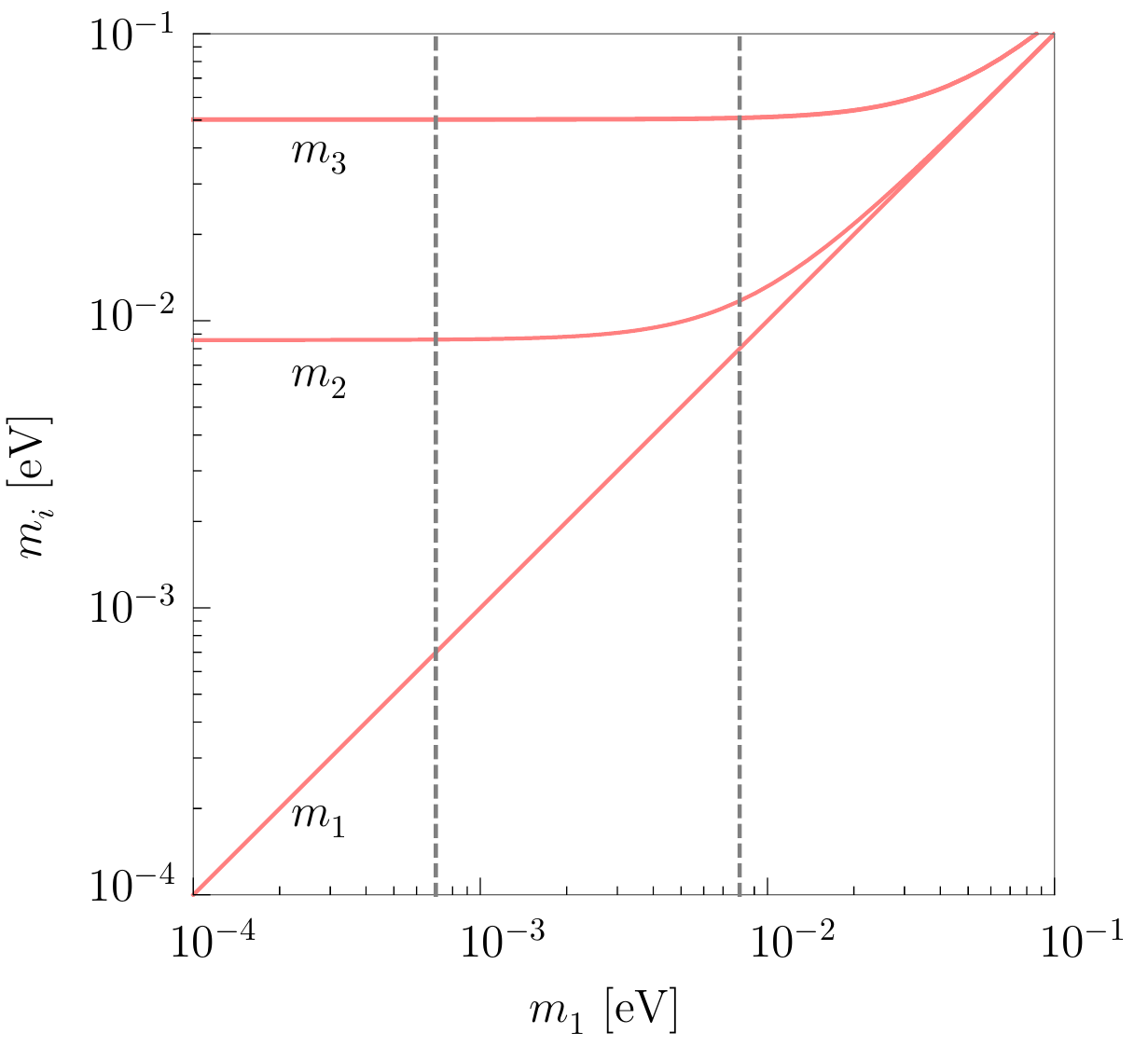}
\end{center}
\vspace{-0.5cm}
\caption{Illustration for the absolute neutrino masses $m^{}_i$ (for $i = 1, 2, 3$) as implied by the sensitivity of $|m^{}_{\beta\beta}| \approx 1~{\rm meV}$ for future $0\nu\beta\beta$ decay experiments, where the allowed range of $m^{}_1 \in [0.7, 8]~{\rm meV}$ is lying between two vertical dashed lines.}
\label{fig:mi}
\end{figure}

Since the lightest neutrino mass $m^{}_1$ can be constrained to a narrow range of $[0.7, 8]~{\rm meV}$ for $|m^{}_{\beta\beta}| \lesssim 1~{\rm meV}$, it is interesting in the first place to see how well the full neutrino mass spectrum can be determined. In Fig.~\ref{fig:mi}, the absolute neutrino masses $m^{}_i$ (for $i = 1, 2, 3$) have been plotted as three red curves against the lightest neutrino mass $m^{}_1$. The requirement for $|m^{}_{\beta\beta}| \lesssim 1~{\rm meV}$ leads to
\begin{eqnarray}\label{eq:mi}
m^{}_1 \in [0.7, 8]~{\rm meV} \; , \quad m^{}_2 \in [8.6, 11.7]~{\rm meV} \; , \quad m^{}_3 \in [50.3, 50.9]~{\rm meV} \; ,
\end{eqnarray}
where the precisions on neutrino oscillation parameters after the JUNO measurements are used. The allowed range of the lightest neutrino mass $m^{}_1 \in [0.7, 8]~{\rm meV}$ is lying between two vertical dashed lines in Fig.~\ref{fig:mi}, where one can accordingly find out the allowed ranges of $m^{}_2$ and $m^{}_3$ on the vertical axis by following the intersecting points between the red curves and the dashed lines. From Eq.~(\ref{eq:mi}), one immediately observes that $m^{}_3$ is already determined with an excellent precision, while the uncertainties on $m^{}_2$ and $m^{}_1$ are also acceptable. It is worthwhile to stress that the complete determination of neutrino mass spectrum will be very suggestive for the model building of neutrino masses.

\begin{figure}[t!]
	\begin{center}
\vspace{-0.4cm}
		\subfigure{
			\includegraphics[width=0.44\textwidth]{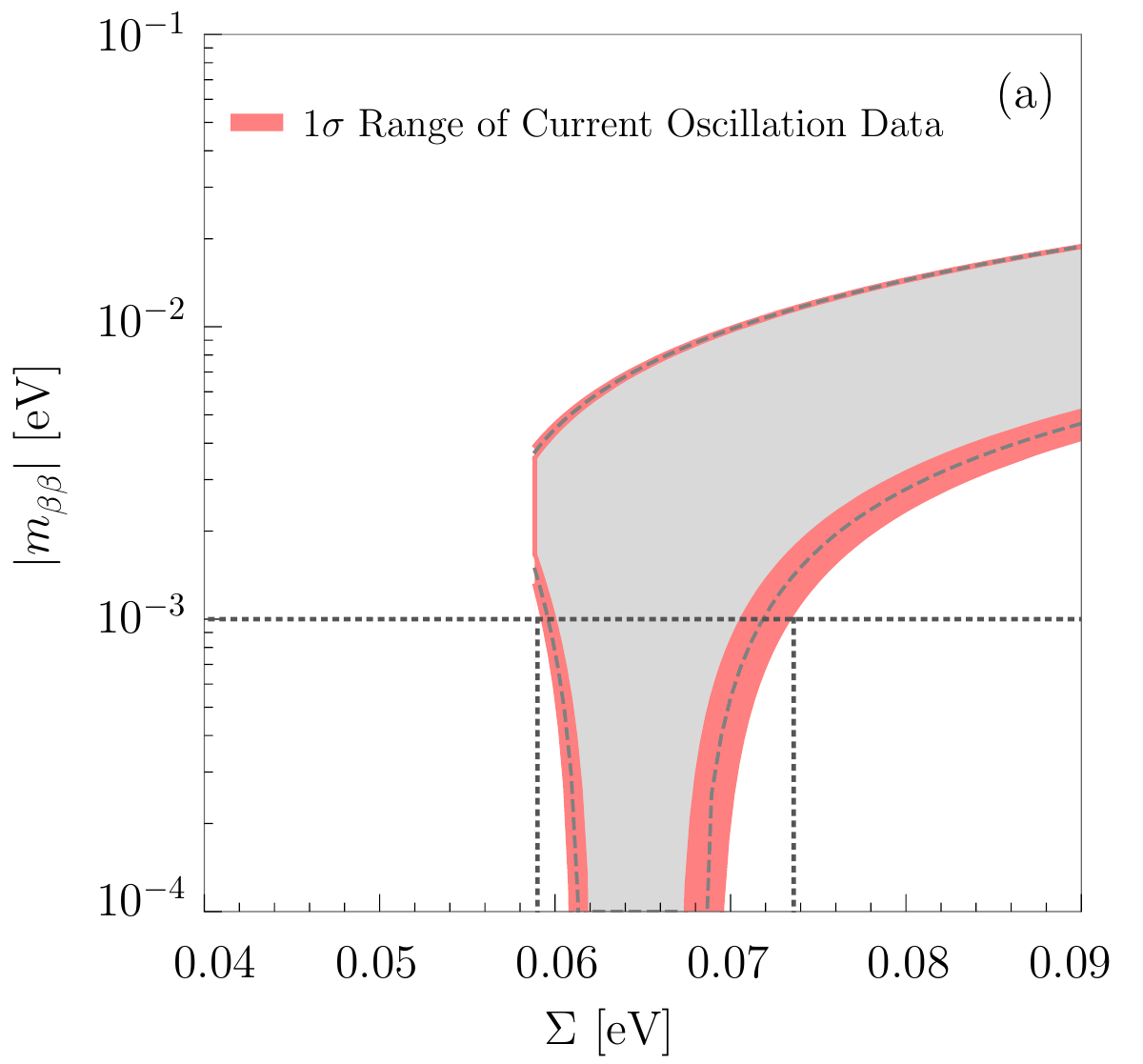} }
		\subfigure{
			\includegraphics[width=0.44\textwidth]{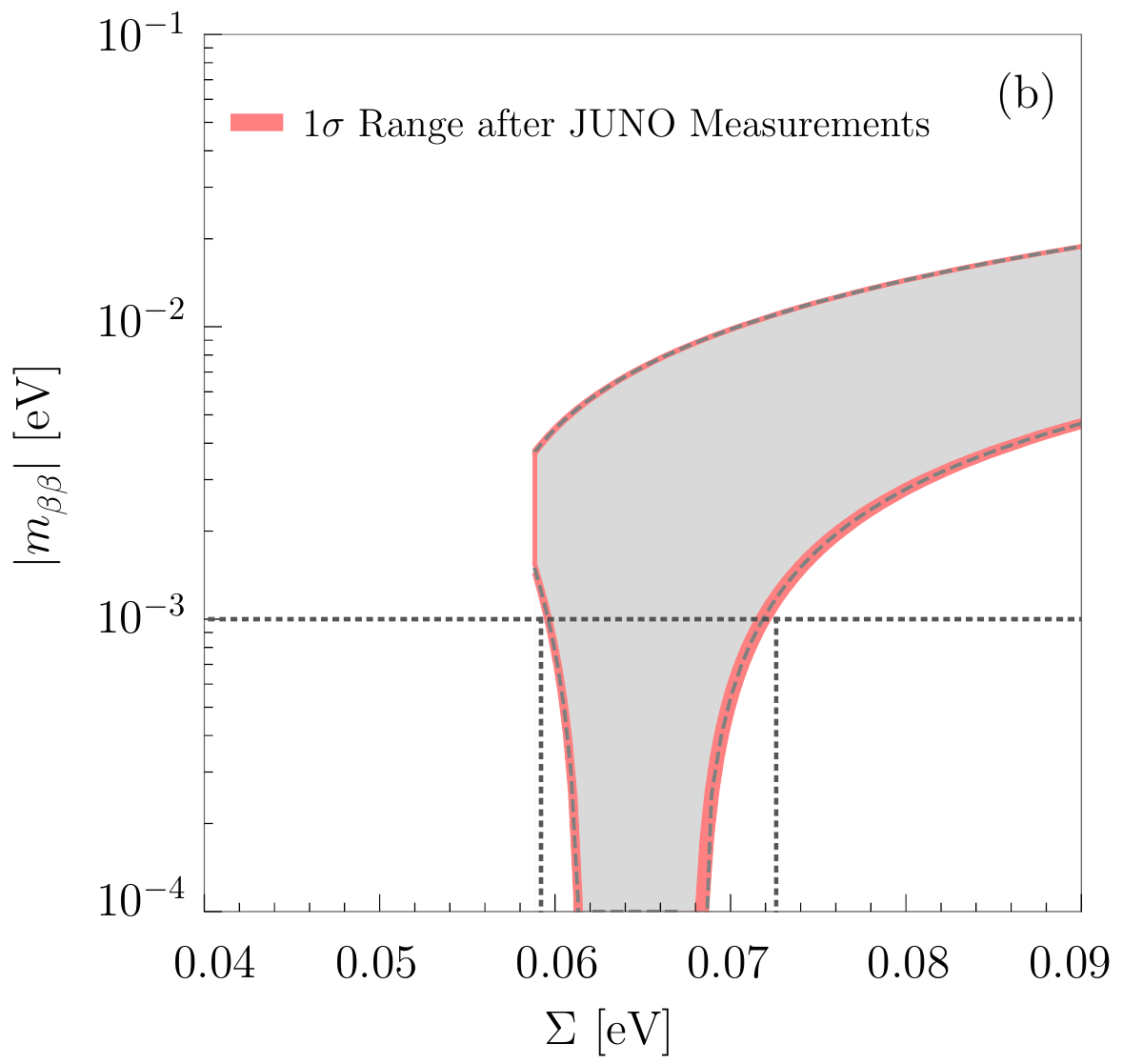} }
		\subfigure{
			\includegraphics[width=0.44\textwidth]{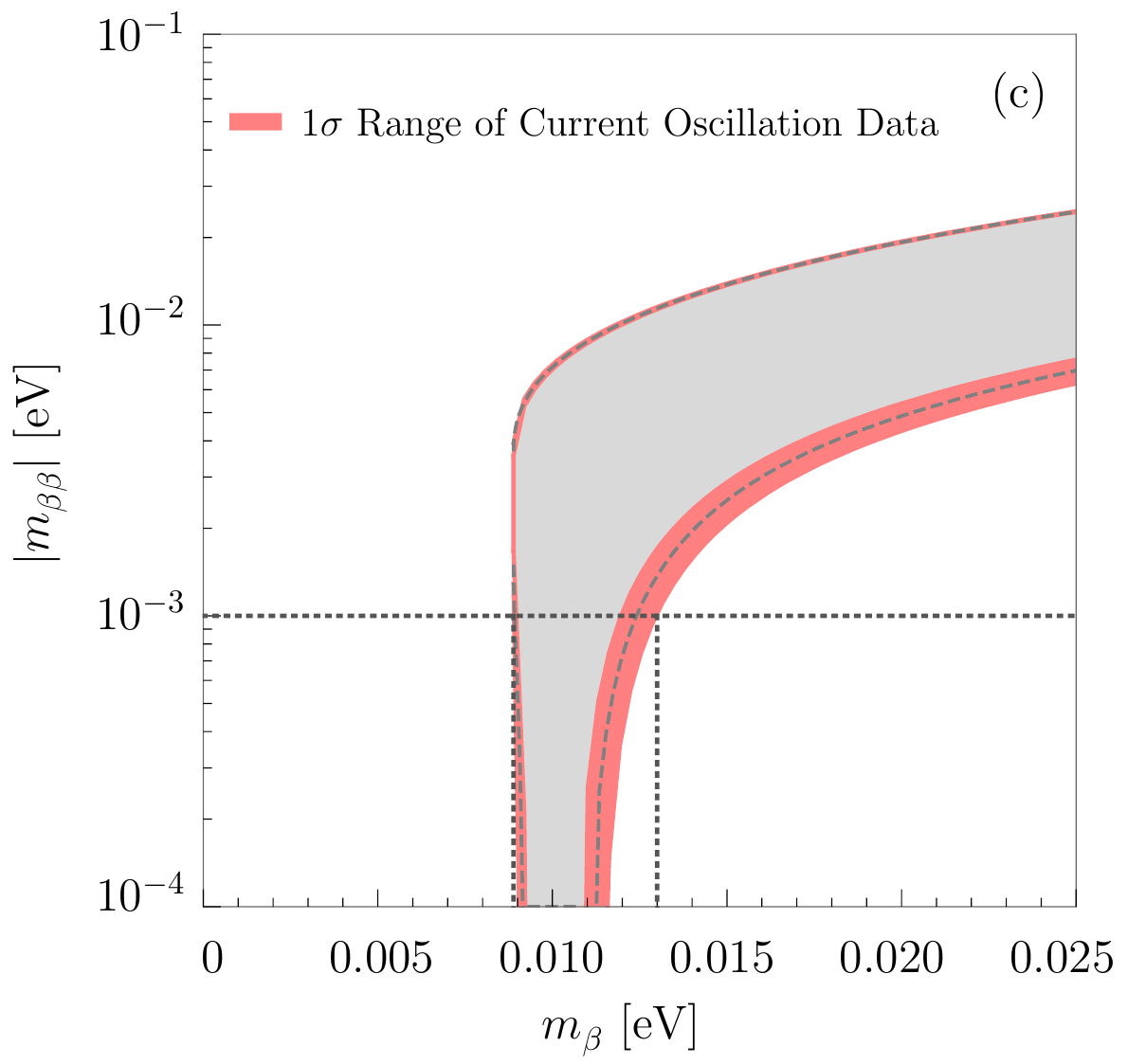} }
		\subfigure{
			\includegraphics[width=0.44\textwidth]{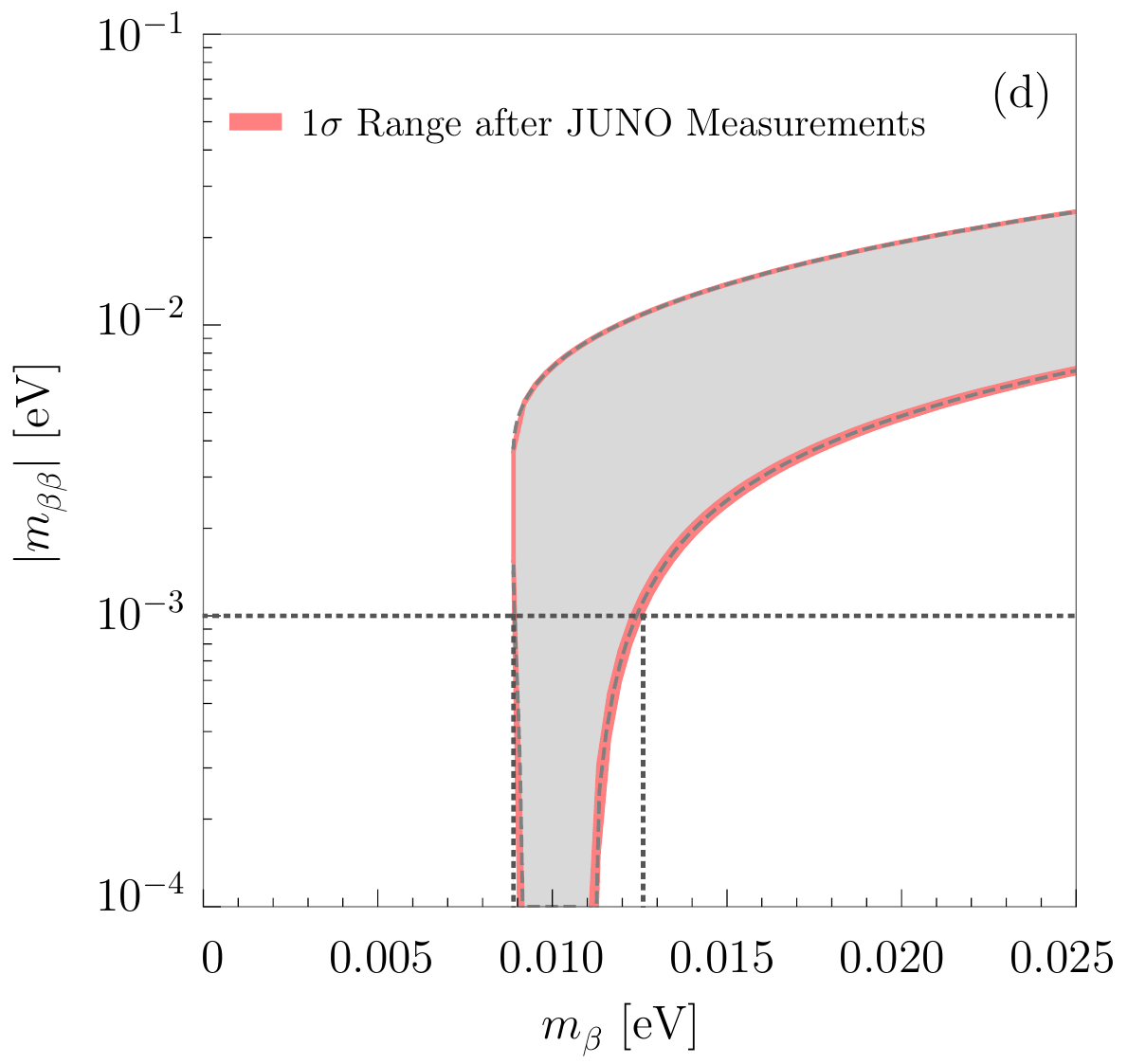} }
		\subfigure{
			\includegraphics[width=0.44\textwidth]{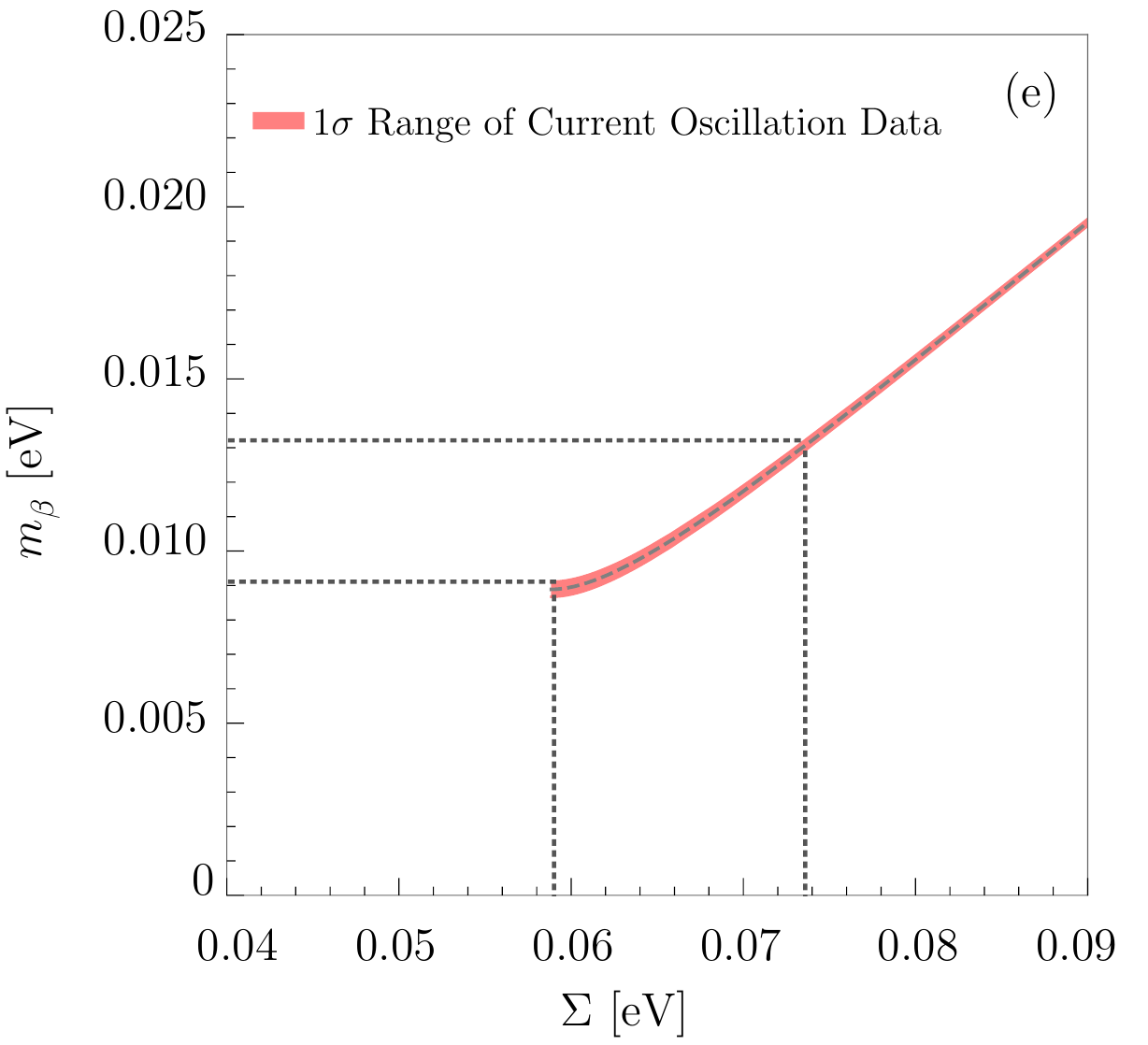} }
		\subfigure{
			\includegraphics[width=0.44\textwidth]{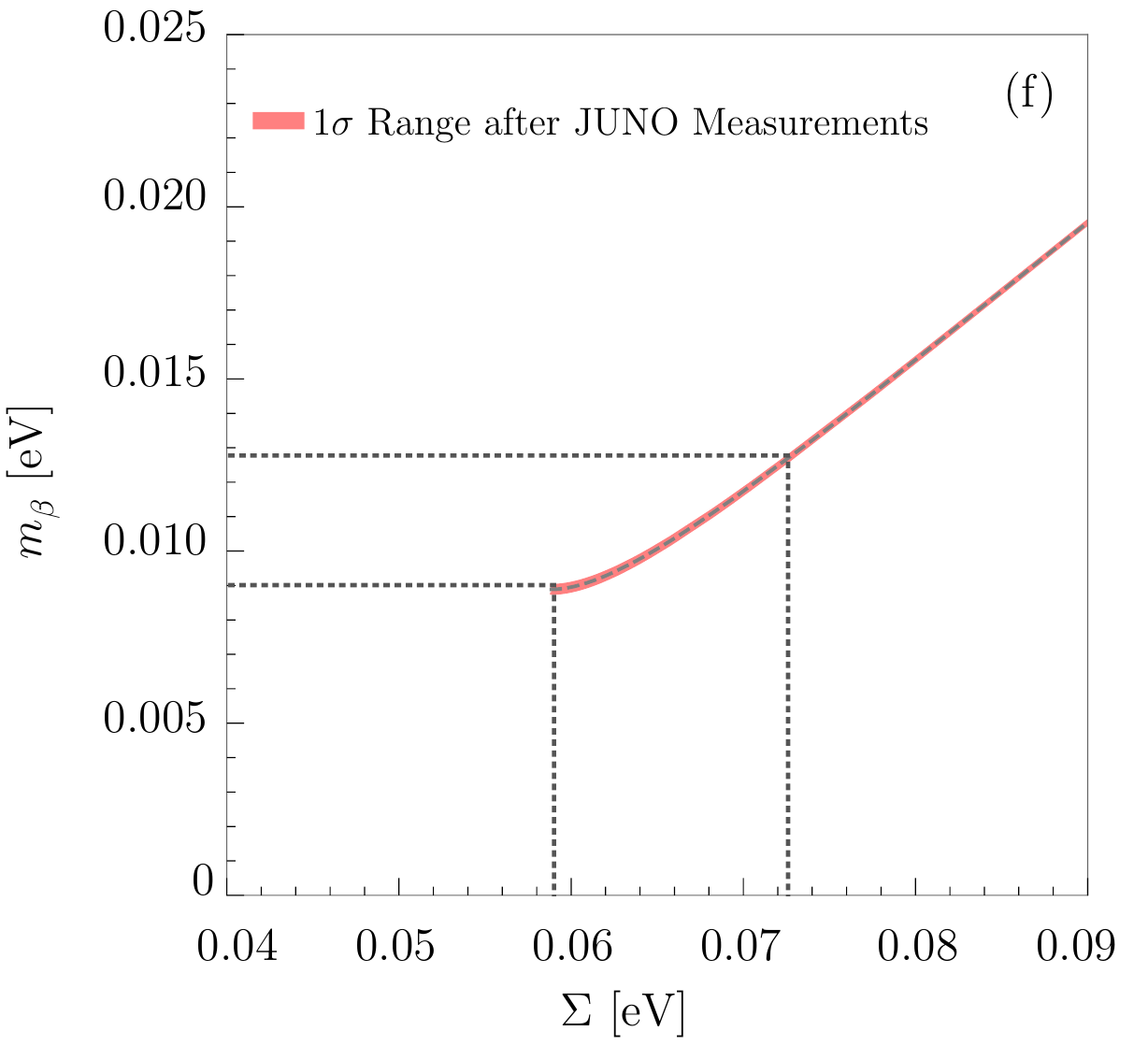} }
	\end{center}
	\vspace{-0.5cm}
	\caption{The allowed regions of three observables $|m^{}_{\beta\beta}|$, $m^{}_\beta$ and $\Sigma$, where the global-fit results of current neutrino oscillation data have been used for three plots in the left column while the JUNO measurements are taken into account for those in the right column.}
	\label{fig:mbsigma}
\end{figure}

Then we explore the implications for the effective neutrino mass $m^{}_\beta$ in beta decays and the sum of three neutrino masses $\Sigma$. For this purpose, we have depicted the allowed regions of three observables $|m^{}_{\beta\beta}|$ from $0\nu\beta\beta$ decays, $m^{}_\beta$ from beta decays, and $\Sigma$ from cosmological observations in Fig.~\ref{fig:mbsigma}. For three plots in the left column, the global-fit results of neutrino oscillation parameters in Eq.~(\ref{eq:data}) have been input, while for those in right column we have considered the projected precisions after the JUNO measurements. The dashed curves as the boundaries of the gray shaded areas have been obtained by using the best-fit values, and the red bands are caused by the $1\sigma$ uncertainties of the input parameters. The remarkable improvements after the JUNO measurements are transparent. For $|m^{}_{\beta\beta}| = 1~{\rm meV}$, which has been shown as the horizontal dashed line in the plots in the first two rows of Fig.~\ref{fig:mbsigma}, one can immediately extract the corresponding ranges of $m^{}_\beta$ and $\Sigma$. For instance, assuming the $1\sigma$ uncertainties of neutrino oscillation parameters after the JUNO experiment, we can obtain
\begin{eqnarray}\label{eq:mbsigma}
8.9~{\rm meV} \leq m^{}_\beta \leq 12.6~{\rm meV} \; , \quad 59.2~{\rm meV} \leq \Sigma \leq 72.6~{\rm meV} \; ,
\end{eqnarray}
which is respectively below the forecasted sensitivity $m^{}_\beta \lesssim 40~{\rm meV}$ of the forthcoming beta-decay experiment~\cite{Esfahani:2017dmu} and that $\Sigma \lesssim 80~{\rm meV}$ of future observations of cosmic microwave background~\cite{Dvorkin:2019jgs}.

As the Majorana CP phases are only involved in the $0\nu\beta\beta$ decays, it is impossible to completely pin down all three unknown parameters $m^{}_1$, $\rho$ and $\sigma$ from such a single type of observations. If the effective neutrino mass $m^{}_\beta$ can be precisely measured in beta-decay experiments, one will be able to solve the lightest neutrino mass $m^{}_1$ with the help of neutrino oscillation data, namely,
\begin{eqnarray}
m^2_1 = m^2_\beta - \Delta m^2_{21} \cos^2\theta^{}_{13} \sin^2 \theta^{}_{12} - \Delta m^2_{31} \sin^2 \theta^{}_{13} \; , \label{eq:mbeta}
\end{eqnarray}
and then to predict $\Sigma = m^{}_1 + \sqrt{m^2_1 + \Delta m^2_{21}} + \sqrt{m^2_1 + \Delta m^2_{31}}$. Moreover, inserting the determined neutrino masses $m^{}_1$, $m^{}_2 = \sqrt{m^2_1 + \Delta m^2_{21}}$ and $m^{}_3 = \sqrt{m^2_1 + \Delta m^2_{31}}$ into the effective neutrino mass $|m^{}_{\beta\beta}|$, we can have a good opportunity to probe the Majorana CP phases $\rho$ and $\sigma$ by improving the sensitivity of $0\nu\beta\beta$ decay experiments to $|m^{}_{\beta\beta}| \approx 1~{\rm meV}$. But this seems to be not the case.

For this reason, we take a different strategy to pin down neutrino masses in the assumption that both $|m^{}_{\beta\beta}|$ and $\Sigma$ can be better measured in the foreseeable future. Given the information on the Majorana CP phases, it is straightforward to establish the following relations
\begin{eqnarray}
\left(\begin{matrix} {\rm Re} \left(U^2_{e1}\right) & {\rm Re} \left(U^2_{e2}\right) & {\rm Re} \left(U^2_{e3}\right) \cr {\rm Im} \left(U^2_{e1}\right) & {\rm Im} \left(U^2_{e2}\right) & {\rm Im} \left(U^2_{e3}\right) \cr 1 & 1 & 1 \end{matrix}\right) \cdot \left(\begin{matrix} m^{}_1 \cr m^{}_2 \cr m^{}_3 \end{matrix}\right) = \left(\begin{matrix} {\rm Re}\left(m^{}_{\beta\beta}\right) \cr {\rm Im}\left(m^{}_{\beta\beta}\right) \cr \Sigma \end{matrix}\right) \; ,
\label{eq:linear}
\end{eqnarray}
where the flavor mixing matrix elements are given by $U^{}_{e1} = \cos \theta^{}_{13} \cos \theta^{}_{12} e^{{\rm i}\rho/2}$, $U^{}_{e2} = \cos \theta^{}_{13} \cos \theta^{}_{12}$ and $U^{}_{e3} = \sin\theta^{}_{13} e^{{\rm i}\sigma/2}$ in accordance with the parametrization adopted in Eq.~(\ref{eq:mbb}). After directly solving Eq.~(\ref{eq:linear}) for neutrino masses, one arrives at
\begin{eqnarray}
m^{}_1 &=& \frac{{\rm Im}\left[(m^{}_{\beta\beta} - \Sigma \cdot U^2_{e2})^* \cdot (U^2_{e2} - U^2_{e3})\right]}{{\rm Im}\left[(U^2_{e3} - U^2_{e1})^* \cdot (U^2_{e1} - U^2_{e2})\right]} \; , \nonumber \\
m^{}_2 &=& \frac{{\rm Im}\left[(m^{}_{\beta\beta} - \Sigma \cdot U^2_{e3})^* \cdot (U^2_{e3} - U^2_{e1})\right]}{{\rm Im}\left[(U^2_{e3} - U^2_{e1})^* \cdot (U^2_{e1} - U^2_{e2})\right]} \; , \label{eq:numass} \\
m^{}_3 &=& \frac{{\rm Im}\left[(m^{}_{\beta\beta} - \Sigma \cdot U^2_{e1})^* \cdot (U^2_{e1} - U^2_{e2})\right]}{{\rm Im}\left[(U^2_{e3} - U^2_{e1})^* \cdot (U^2_{e1} - U^2_{e2})\right]} \; , \nonumber
\end{eqnarray}
from which we can easily verify that $m^{}_1 + m^{}_2 + m^{}_3 = \Sigma$ holds as it should do. Notice that the denominator ${\rm Im}\left[(U^2_{e3} - U^2_{e1})^* \cdot (U^2_{e1} - U^2_{e2})\right]$ is supposed to be nonzero in Eq.~(\ref{eq:numass}). If it is zero for some specific values of $\rho$ and $\sigma$, then we can extract neutrino masses just from the effective mass $|m^{}_{\beta\beta}|$ and neutrino oscillation parameters.

As an application of Eq.~(\ref{eq:numass}), let us consider the special scenario in which $|m^{}_{\beta\beta}| = 0$ is realized. In this particular case, both $\rho$ and $\sigma$ will be fixed for the complete cancellation in $|m^{}_{\beta\beta}|$ to happen~\cite{XZ}. Then, we can find three neutrino masses
\begin{eqnarray}
m^{}_1 &=&  \frac{+ \Sigma \cdot |U^{}_{e2}|^2 |U^{}_{e3}|^2 \sin\sigma}{|U^{}_{e2}|^2 |U^{}_{e3}|^2 \sin\sigma + |U^{}_{e1}|^2 |U^{}_{e3}|^2 \sin(\rho - \sigma) - |U^{}_{e1}|^2 |U^{}_{e2}|^2 \sin\rho} \; , \nonumber \\
m^{}_2 &=&  \frac{+ \Sigma \cdot |U^{}_{e1}|^2 |U^{}_{e3}|^2 \sin(\rho - \sigma)}{|U^{}_{e2}|^2 |U^{}_{e3}|^2 \sin\sigma + |U^{}_{e1}|^2 |U^{}_{e3}|^2 \sin(\rho - \sigma) - |U^{}_{e1}|^2 |U^{}_{e2}|^2 \sin\rho} \; , \label{eq:numasssol} \\
m^{}_3 &=&  \frac{- \Sigma \cdot |U^{}_{e1}|^2 |U^{}_{e2}|^2 \sin\rho}{|U^{}_{e2}|^2 |U^{}_{e3}|^2 \sin\sigma + |U^{}_{e1}|^2 |U^{}_{e3}|^2 \sin(\rho - \sigma) - |U^{}_{e1}|^2 |U^{}_{e2}|^2 \sin\rho} \; , \nonumber
\end{eqnarray}
where one can observe that neutrino masses are evidently proportional to $\Sigma$. Certainly, the same results can also be obtained by requiring the real and imaginary parts of $m^{}_{\beta\beta}$ to be vanishing and then deriving two independent neutrino mass ratios~\cite{Xing:2003jf, Xing:2003ez}.

\section{Concluding Remarks}\label{sec:conc}

In the present work, we have explained why it is important for the future $0\nu\beta\beta$ decay experiments to reach the sensitivity of $|m^{}_{\beta\beta}| \approx 1~{\rm meV}$. First of all, with such a high sensitivity, one can place restrictive constraints on the lightest neutrino mass $0.7~{\rm meV} \leq m^{}_1 \leq 8~{\rm meV}$ and one of the Majorana CP phases $130^\circ \leq \rho \leq 230^\circ$. Second, these constraints further imply that neutrino mass spectrum is almost fixed, namely, $m^{}_1 \in [0.7, 8]~{\rm meV}$, $m^{}_2 \in [8.6, 11.7]~{\rm meV}$ and $m^{}_3 \in [50.3, 50.9]~{\rm meV}$, the effective neutrino mass for beta decays should be lying in the range $8.9~{\rm meV} \leq m^{}_{\beta} \leq 12.6~{\rm meV}$ and the sum of three neutrino masses must be $59.2~{\rm meV} \leq \Sigma \leq 72.6~{\rm meV}$. These limits are lying below the forecasted sensitivities from the next-generation beta-decay experiments~\cite{Wolf:2008hf, Esfahani:2017dmu} and the future cosmological observations~\cite{Dvorkin:2019jgs}.

Among all the current $0\nu\beta\beta$ decay experiments in operation~\cite{KamLAND-Zen:2016pfg, Auger:2012ar, Albert:2014awa, Albert:2017owj, Agostini:2017iyd, Agostini:2018tnm, Alduino:2017ehq}, the KamLAND-Zen collaboration has reported the best sensitivity $|m^{}_{\beta\beta}| < (61\cdots 165)~{\rm meV}$ depending on the NME for the $0\nu\beta\beta$ decays of $^{136}{\rm Xe}$. The next-generation experiments aim for $|m^{}_{\beta\beta}| \approx 10~{\rm meV}$, which has been set up as it is the lower boundary of $|m^{}_{\beta\beta}|$ in the IO case~\cite{Dolinski:2019nrj, Bilenky:2014uka}. Therefore, an urgent question is whether it is realistic to reach $|m^{}_{\beta\beta}| \approx 1~{\rm meV}$ in the near future. The studies in Ref.~\cite{Zhao:2016brs} have demonstrated that if the JUNO-LS detector is upgraded with $^{136}{\rm Xe}$-loading in future, a sensitivity of $|m^{}_{\beta\beta}| \approx 5~{\rm meV}$ is achievable when the most optimistic value of the NME is taken. In order to improve the sensitivity to $|m^{}_{\beta\beta}| \approx 1~{\rm meV}$, one has to remarkably increase the target mass and reduce the backgrounds. Although the radioactive and cosmogenic backgrounds can be rejected by severe radiopurity control and perfect muon veto strategies, the irreducible backgrounds arising from recoiled electrons due to their elastic scattering with solar $^8{\rm B}$ neutrinos and the two-neutrino-emitting double-beta decays could be a serious problem. Nevertheless, inspired by the work in Ref.~\cite{Zhao:2016brs}, it is promising to achieve $|m^{}_{\beta\beta}| \approx 1~{\rm meV}$ by developing the $^{130}{\rm Te}$-loaded LS to reach a sufficiently large target mass and advancing the powerful techniques for background reduction.

Strictly speaking, one should perform a statistical analysis of the experimental sensitivities to neutrino masses and the Majorana CP phases~\cite{Zhang:2015kaa, Agostini:2017jim, Caldwell:2017mqu, Huang:2019qvq}. However, we have found that the final results are quite consistent with the simple analysis in the present paper. We believe that our analysis is very suggestive for setting up the future program for $0\nu\beta\beta$ decay experiments. If the sensitivity of $|m^{}_{\beta\beta}| \approx 1~{\rm meV}$ is ultimately realized, the determination of absolute neutrino masses and the constraints on one of two Majorana CP phases are possible, which cannot be accessible in other types of feasible neutrino experiments. To achieve this goal, we may have to make great efforts in increasing the target mass and reducing the background by two orders of magnitude compared to the present design of next-generation $0\nu\beta\beta$ decay experiments.

\section*{Acknowledgements}

This work was supported in part by the National Key R\&D Program of China under Grant No. 2018YFA0404100, by the Strategic Priority Research Program of the Chinese Academy of Sciences under Grant No. XDA10010100, by the National Natural Science Foundation of China under Grant No.~11605081, No.~11775231, No.~11775232, No.~11835013 and No.~11820101005, and by the CAS Center for Excellence in Particle Physics.

\end{document}